\documentclass[amsmath,aps,superscriptaddress]{revtex4}

 \usepackage{epsf}
 \usepackage{graphicx}    

\begin{document}

 \newcommand{\be}[1]{\begin{equation}\label{#1}}
 \newcommand{\ee}{\end{equation}}
 \newcommand{\bea}{\begin{eqnarray}}
 \newcommand{\eea}{\end{eqnarray}}
 \def\disp{\displaystyle}

 \def\gsim{ \lower .75ex \hbox{$\sim$} \llap{\raise .27ex \hbox{$>$}} }
 \def\lsim{ \lower .75ex \hbox{$\sim$} \llap{\raise .27ex \hbox{$<$}} }



 \title{Agegraphic dark energy as a quintessence}

\author{Jingfei Zhang}
\affiliation{School of Physics and Optoelectronic Technology,
Dalian University of Technology, Dalian 116024, People's Republic
of China}
\author{Xin Zhang}
\affiliation{Kavli Institute for Theoretical Physics China,
Institute of Theoretical Physics, Chinese Academy of Sciences
(KITPC/ITP-CAS), P.O.Box 2735, Beijing 100080, People's Republic of
China}
\author{Hongya Liu}
\affiliation{School of Physics and Optoelectronic Technology,
Dalian University of Technology, Dalian 116024, People's Republic
of China}

\begin{abstract}
Recently, a dark energy model characterized by the age of the
universe, dubbed ``agegraphic dark energy'', was proposed by Cai. In
this paper, a connection between the quintessence scalar-field and
the agegraphic dark energy is established, and accordingly, the
potential of the agegraphic quintessence field is constructed.
\end{abstract}


 \maketitle


 \renewcommand{\baselinestretch}{1.6}



\section{Introduction}\label{sec1}
Many cosmological experiments, such as observations of large scale
structure (LSS)~\cite{LSS}, searches for type Ia supernovae
 (SNIa)~\cite{SN} and measurements of the cosmic microwave
background (CMB) anisotropy~\cite{CMB}, all suggest that our
universe is experiencing an accelerating expansion at the present
time. Following the standard Friedmann-Robertson-Walker (FRW)
cosmology such an expansion implies the existence of a mysterious
dominant component, dark energy, with large enough negative
pressure. Although we can affirm that the ultimate fate of the
universe is determined by the feature of dark energy, the nature of
dark energy as well as its cosmological origin remain enigmatic (for
reviews, see e.g. \cite{Weinberg88cp,Carroll00fy,Peebles02gy}). It
is fair to say that disclosing the nature of dark energy is one of
the central problems in the research of both cosmology and
theoretical physics at present. Researchers proposed many candidates
to interpret or describe the properties of dark energy. The simplest
candidate for dark energy is the famous cosmological constant which
has the equation of state $w=-1$. However, the cosmological constant
scenario, is plagued with the so-called ``fine-tuning problem'' and
``coincidence problem''~\cite{lamada}. Theorists have made lots of
efforts to try to resolve the cosmological constant problem, but all
these efforts seem to be unsuccessful. Of course the theoretical
consideration is still in process and has made some progresses.

An alternative proposal for dark energy is the dynamical dark energy
scenario which is often realized by scalar field mechanism. It
suggests that the energy form with negative pressure is provided by
a scalar field evolving down a proper potential.  A famous example
of scalar-field dark energy is the so-called ``quintessence''
\cite{quintessence}. Provided that the evolution of the field is
slow enough, the kinetic energy density is less than the potential
energy density, giving rise to the negative pressure responsible to
the cosmic acceleration. So far, besides quintessence, a host of
scalar-field dark energy models have been studied such as
$K$-essence \cite{kessence}, hessence \cite{hessence}, tachyon
\cite{tachyon}, phantom \cite{phantom}, ghost condensate
\cite{ghost} and quintom \cite{quintom}, and so forth. But we should
note that the mainstream viewpoint regards the scalar field dark
energy models as an effective description of an underlying theory of
dark energy. In addition, other proposals on dark energy include
interacting dark energy models \cite{intde}, braneworld models
\cite{brane}, and Chaplygin gas models \cite{cg}, etc..

However, we still can make some attempts to probe the nature of dark
energy according to some principles of quantum gravity although a
complete theory of quantum gravity is not available today. The
holographic dark energy model \cite{holo1,holo2,holo3,holo4} is just
an appropriate and interesting example, which is constructed in the
light of the holographic principle of quantum gravity theory. That
is to say, the holographic dark energy model possesses some
significant features of an underlying theory of dark energy. More
recently, a new dark energy model, dubbed agegraphic dark energy,
has been proposed~\cite{rgcai}, which takes into account the
uncertainty relation of quantum mechanics together with the
gravitational effect in general relativity.

 In the general
relativity, one can measure the spacetime without
 any limit of accuracy. However, in the quantum mechanics, the
 well-known Heisenberg uncertainty relation puts a limit of
 accuracy in these measurements. Following the line of quantum
 fluctuations of spacetime, K\'{a}rolyh\'{a}zy and his
 collaborators~\cite{r1} (see also~\cite{r2}) made an interesting
 observation concerning the distance measurement for Minkowski
 spacetime through a light-clock {\it Gedanken experiment}, namely,
 the distance $t$ in Minkowski spacetime cannot be known to a better
 accuracy than
 \begin{equation}
\delta t=\lambda t_p^{2/3}t^{1/3}~,\label{eq1}
\end{equation}
 where $\lambda$ is a dimensionless constant of order unity. We use
 the units $\hbar=c=k_B=1$ throughout this paper. Thus, one can use
 the terms like length and time interchangeably, whereas
 $l_p=t_p=1/m_p$ with $l_p$, $t_p$ and $m_p$ being the reduced
 Planck length, time and mass respectively.

The K\'{a}rolyh\'{a}zy relation~(\ref{eq1}) together with the
 time-energy uncertainty relation enables one to estimate a quantum
 energy density of the metric fluctuations of Minkowski
 spacetime~\cite{r3,r2}. Following~\cite{r3,r2}, with respect to
 the Eq.~(\ref{eq1}) a length scale $t$ can be known with a maximum
 precision $\delta t$ determining thereby a minimal detectable cell
 $\delta t^3\sim t_p^2 t$ over a spatial region $t^3$. Such a cell
 represents a minimal detectable unit of spacetime over a given
 length scale $t$. If the age of the Minkowski spacetime is $t$,
 then over a spatial region with linear size $t$ (determining the
 maximal observable patch) there exists a minimal cell $\delta t^3$
 the energy of which due to time-energy uncertainty relation can not
 be smaller than~\cite{r3,r2}
\begin{equation}
 E_{\delta t^3}\sim t^{-1}~.\label{eq2}
 \end{equation}
 Therefore, the energy density of metric fluctuations of
 Minkowski spacetime is given by~\cite{r3,r2}
 \begin{equation}
 \rho_q\sim\frac{E_{\delta t^3}}{\delta t^3}\sim
 \frac{1}{t_p^2 t^2}\sim\frac{m_p^2}{t^2}~.\label{eq3}
 \end{equation}

In~\cite{r3} (see also~\cite{rgcai}), it is noticed that the
 K\'{a}rolyh\'{a}zy relation~(\ref{eq1}) naturally obeys the
 holographic black hole entropy bound. In fact, the holographic
 dark energy~\cite{holo1} also stems from the the idea of holographic
 black hole entropy bound. It is worth noting that
 the form of energy density Eq.~(\ref{eq3}) is similar to the one
 of holographic dark energy~\cite{holo1}, i.e.,
 $\rho_\Lambda\sim l_p^{-2}l^{-2}$. The similarity between
 $\rho_q$ and $\rho_\Lambda$ might reveal some universal features
 of quantum gravity, although they arise from different ways.
But, the agegraphic dark energy model avoids the causality problem
which exists in the holographic dark energy models. For extensive
studies on agegraphic dark energy,
see~\cite{Wei:2007ut,Wei:2007zs,Wu:2007wu,Wei:2007ty,Zhang:2007ps,Wei:2007xu}.

By far, hundreds of dark energy models have been constructed.
However, it is hard to say that we are close to the Grail of
revealing the nature of dark energy (that almost means we can
understand the quantum gravity well). Although going along the
holographic principle of quantum gravity may provide a hopeful way
towards the aim, it is hard to believe that the physical foundation
of agegraphic dark energy is convincing enough. Actually, it is fair
to say that almost all dynamical dark energy models are settled at
the phenomenological level, neither holographic dark energy model
nor agegraphic dark energy model is exception.

Though, under such circumstances, the models of holographic and
agegraphic dark energy, to some extent, still have some advantage
comparing to other dynamical dark energy models because at least
they are built according to some fundamental principle---holographic
principle---in quantum gravity. We thus may as well view that this
class of models possesses some features of an underlying theory of
dark energy. Now, we are interested in that if we assume the
agegraphic dark energy scenario as the underlying theory of dark
energy, how the low-energy effective scalar-field model can be used
to describe it. In this direction, we can establish a correspondence
between the agegraphic dark energy and quintessence scalar field,
and describe agegraphic dark energy in this case effectively by
making use of quintessence. We refer to this case as ``agegraphic
quintessence''. The quintessence potential $V(\phi)$ can be
reconstructed from supernova observational data
\cite{Saini:1999ba,simplescalar}. In addition, from some specific
parametrization forms of the equation of state $w(z)$, one can also
reconstruct the quintessence potential $V(\phi)$ \cite{Guo:2005at}.
The reconstruction method can also be generalized to scalar-tensor
theories \cite{scalartensor}, $f(R)$ gravity \cite{frgrav}, a dark
energy fluid with viscosity terms \cite{viscosity}, and also the
generalized ghost condensate model \cite{Zhang:2006em}. For a
reconstruction program for a very general scalar-field Lagrangian
density see \cite{general}. For the scalar-field effective
description of the holographic dark energy see \cite{holoscalar}. In
this paper, we shall reconstruct the quintessence potential and the
dynamics of the scalar field in the light of the agegraphic dark
energy.

In the next section, we briefly review the original agegraphic dark
energy model proposed in~\cite{rgcai} and the new agegraphic dark
energy model proposed in~\cite{Wei:2007ty}. In Sec.~\ref{sec3}, we
establish the correspondence between the agegraphic dark energy and
the quintessence. The quintessence potential and the dynamics of the
scalar field are also constructed in the light of the agegraphic
dark energy. Conclusion is given in Sec.~\ref{sec5}.


\section{Agegraphic dark energy models}\label{sec2}

\subsection{Model 1: the original agegraphic dark energy model }\label{sec2a}
Based on the energy density~(\ref{eq3}), a so-called agegraphic
dark
 energy model was proposed in~\cite{rgcai}. There, as the most natural
 choice, the time scale $t$ in Eq.~(\ref{eq3}) is chosen to
 be the age of the universe
 \be{eq4}
 T=\int_0^a\frac{da}{Ha},
 \ee
 where $a$ is the scale factor of our universe; $H\equiv\dot{a}/a$
 is the Hubble parameter; a dot denotes the derivative with respect
 to cosmic time. Thus, the energy density of the agegraphic dark
 energy is given by~\cite{rgcai}
 \be{eq5}
 \rho_q=\frac{3n^2m_p^2}{T^2},
 \ee
 where the numerical factor $3n^2$ is introduced to parameterize
 some uncertainties, such as the species of quantum fields in
 the universe, the effect of curved spacetime (since the energy
 density is derived for Minkowski spacetime), and so on. Obviously,
 since the present age of the universe $T_0\sim H_0^{-1}$ (the
 subscript ``0'' indicates the present value of the corresponding
 quantity; we set $a_0=1$), the present energy density of the
 agegraphic dark energy explicitly meets the observed value
 naturally, provided that the numerical factor $n$ is of order
 unity. In addition, by choosing the age of the universe rather than
 the future event horizon as the length measure, the drawback
 concerning causality in the holographic dark energy model \cite{holo1}
 does not exist in the agegraphic dark energy model~\cite{rgcai}.

If we consider a flat Friedmann-Robertson-Walker~(FRW) universe
 containing agegraphic dark energy and pressureless matter, the
 corresponding Friedmann equation reads
 \be{eq6}
 H^2=\frac{1}{3m_p^2}\left(\rho_m+\rho_q\right).
 \ee
 or equivalently,
\begin{equation}
E(z)\equiv {H(z)\over H_0}=\left(\Omega_{\rm m0}(1+z)^3\over
1-\Omega_{\rm q}\right)^{1/2},\label{Ez}
\end{equation} where $z=(1/a)-1$ is the redshift of the universe.
 It is convenient to introduce the fractional energy densities
 $\Omega_i\equiv\rho_i/(3m_p^2H^2)$ for $i=m$ and $q$. From
 Eq.~(\ref{eq5}), it is easy to find
 \be{eq8}
 \Omega_q=\frac{n^2}{H^2T^2},
 \ee
 whereas $\Omega_m=1-\Omega_q$ from Eq.~(\ref{eq6}). By using
 Eqs.~(\ref{eq5})---(\ref{eq8}) and the energy conservation
 equation $\dot{\rho}_m+3H\rho_m=0$, we obtain the equation
 of motion for $\Omega_q$ as~\cite{rgcai}
 \be{eq9}
 \Omega_q^\prime=\Omega_q\left(1-\Omega_q\right)
 \left(3-\frac{2}{n}\sqrt{\Omega_q}\right),
 \ee
 where a prime denotes the derivative with respect to
 $N\equiv\ln a$. Obviously,
Eq.~(\ref{eq9}) can be rewritten as \be{eq10}
 \frac{d\Omega_q}{dz}=-(1+z)^{-1}\Omega_q\left(1-\Omega_q\right)
 \left(3-\frac{2}{n}\sqrt{\Omega_q}\right).
 \ee
 From the energy conservation
 equation $\dot{\rho}_q+3H(\rho_q+p_q)=0$, as well as
 Eqs.~(\ref{eq5}) and~(\ref{eq8}), it is easy to find that the
 equation-of-state parameter~(EoS) of the agegraphic dark energy
 $w_q\equiv p_q/\rho_q$ is given by~\cite{rgcai}
 \be{eq11}
 w_q=-1+\frac{2}{3n}\sqrt{\Omega_q}.
 \ee

\subsection{Model 2: the new model of agegraphic dark energy}\label{sec2b}
By choosing the time scale to be the conformal time $\eta$, Wei and
Cai proposed a new agegraphic dark energy model~\cite{Wei:2007ty},
the new energy density of the agegraphic dark energy
 reads
 \be{eq12}
 \rho_q=\frac{3n^2m_p^2}{\eta^2},
 \ee
 where the conformal time
 \be{eq13}
 \eta\equiv\int\frac{dt}{a}=\int\frac{da}{a^2H}.
 \ee
 The corresponding fractional energy density is given by
 \be{eq14}
 \Omega_q=\frac{n^2}{H^2\eta^2}.
 \ee

We consider a flat FRW universe containing the new
 agegraphic dark energy and pressureless matter. By using
 Eqs.~(\ref{eq6}), (\ref{eq12}), (\ref{eq13}) and the energy
 conservation equation $\dot{\rho}_m+3H\rho_m=0$, we find that
 the equation of motion for $\Omega_q$ is given by
 \be{eq15}
 \frac{d\Omega_q}{da}=\frac{\Omega_q}{a}\left(1-\Omega_q\right)
 \left(3-\frac{2}{n}\frac{\sqrt{\Omega_q}}{a}\right).
 \ee
Apparently, Eq. (\ref{eq15}) can also be rewritten as \be{eq16}
 \frac{d\Omega_q}{dz}=-\Omega_q\left(1-\Omega_q\right)
 \left(3(1+z)^{-1}-\frac{2}{n}\sqrt{\Omega_q}\right).
 \ee
 From the energy conservation equation
 $\dot{\rho}_q+3H(\rho_q+p_q)=0$, as well as Eqs.~(\ref{eq12})
 and~(\ref{eq14}), it is easy to find that the EoS of the new agegraphic
 dark energy, $w_q\equiv p_q/\rho_q$, is given by
 \be{eq17}
 w_q=-1+\frac{2}{3n}(1+z)\sqrt{\Omega_q}~.
 \ee

In order to compare the behaviors of the two agegraphic dark energy
models, we plot the evolution curves of $w$ in Fig.1 for the two
models with different parameter $n$. It is easy to see that the
behavior of the new agegraphic dark energy is very different from
the original agegraphic energy model. For model 1, the EoS has a
``thawing'' feature. At earlier time where $\Omega_q \to 0$, one has
$w_q \to -1$. Namely, the dark energy behaves like a cosmological
constant at earlier time. At later time where $\Omega_q \to 1$, the
EoS goes to $w_q = -1 +\frac{2}{3n}$. Therefore the fate of our
universe is a forever accelerated expansion with a power-law form in
this dark energy model~\cite{rgcai}. While for model 2, the EoS has
a ``freezing'' feature, i.e., in the late time $w_q\to -1$ when
$a\to\infty$, it mimics a cosmological constant regardless of the
value of $n$.

\begin{figure}[htbp]
\centering $\begin{array}{c@{\hspace{0.2in}}c}
\multicolumn{1}{l}{\mbox{}} &
\multicolumn{1}{l}{\mbox{}} \\
\includegraphics[scale=0.75]{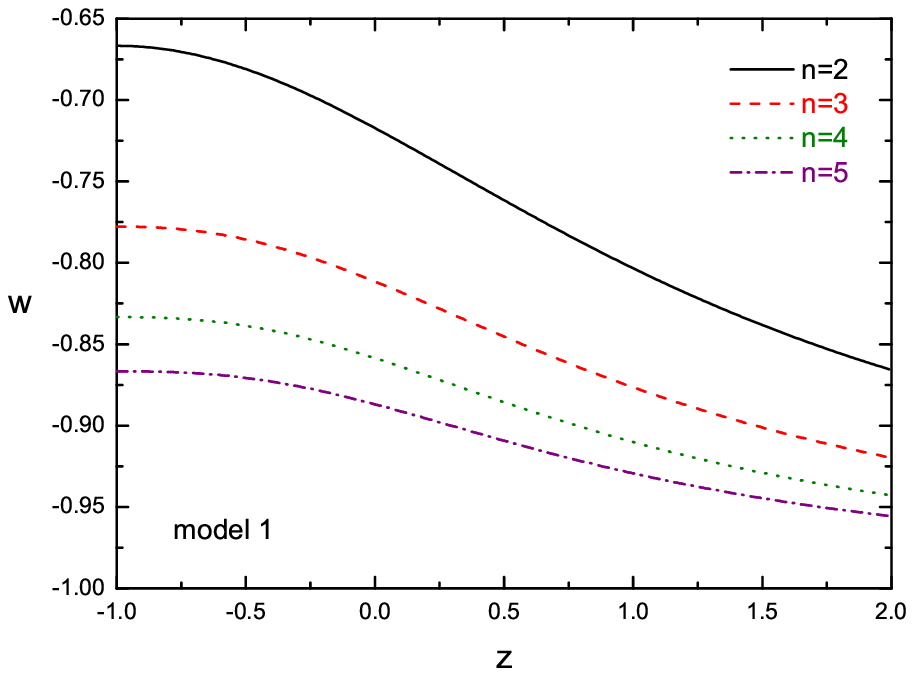} &\includegraphics[scale=0.75]{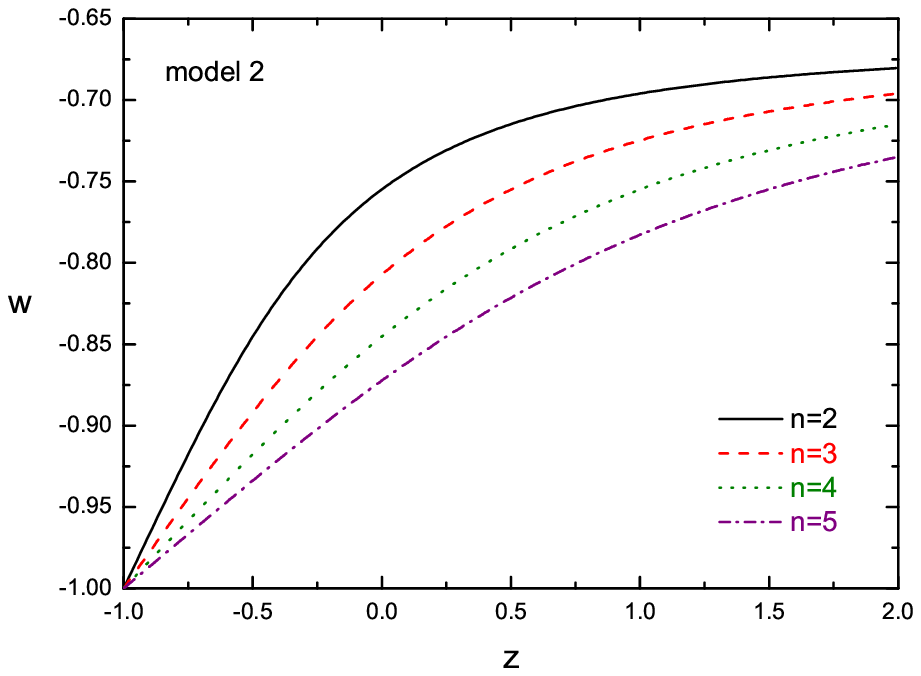} \\
\end{array}$
\caption{The evolution of $w_q$ for model 1 and model 2 with
different model parameter $n$.}
\end{figure}








\section{Reconstructing agegraphic quintessence}\label{sec3}
In this section, we will construct the agegraphic quintessence
models, connecting the quintessence scalar-field with the agegraphic
dark energy.

The quintessence scalar field $\phi$ evolves in its potential
$V(\phi)$ and seeks to roll towards the minimum of the potential,
according to the Klein-Gordon equation
$\ddot{\phi}+3H\dot{\phi}=-dV/d\phi$. The rate of evolution is
driven by the slope of the potential and damped by the cosmic
expansion through the Hubble parameter $H$. The energy density and
pressure are $\rho_\phi=\dot{\phi}^2/2+V$,
$p_\phi=\dot{\phi}^2/2-V$, so that the equation of state of
quintessence $w_\phi=p_\phi/\rho_\phi$ evolves in a region of
$-1<w_\phi<1$. Usually, for making the universe's expansion
accelerate, it should be required that $w_\phi$ must satisfy
$w_\phi<-1/3$. Obviously, both model 1 and model 2 can be described
by the quintessence. We establish a correspondence between the
agegraphic dark energy and quintessence scalar field, and describe
agegraphic dark energy effectively by making use of quintessence. We
refer to this case as ``agegraphic quintessence''.
\begin{figure}[htbp]
\centering $\begin{array}{c@{\hspace{0.2in}}c}
\multicolumn{1}{l}{\mbox{}} &
\multicolumn{1}{l}{\mbox{}} \\
\includegraphics[scale=0.75]{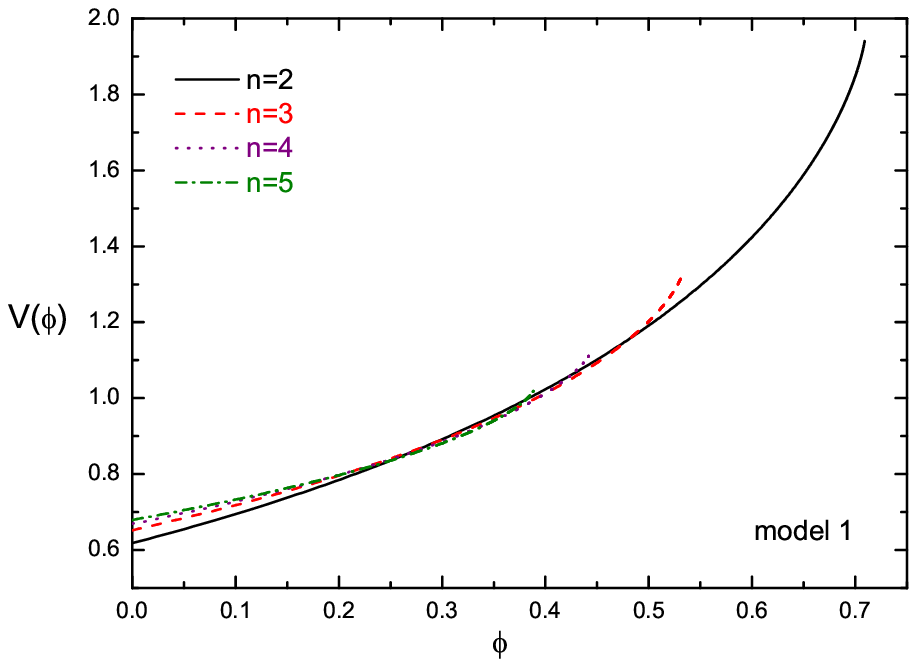} &\includegraphics[scale=0.75]{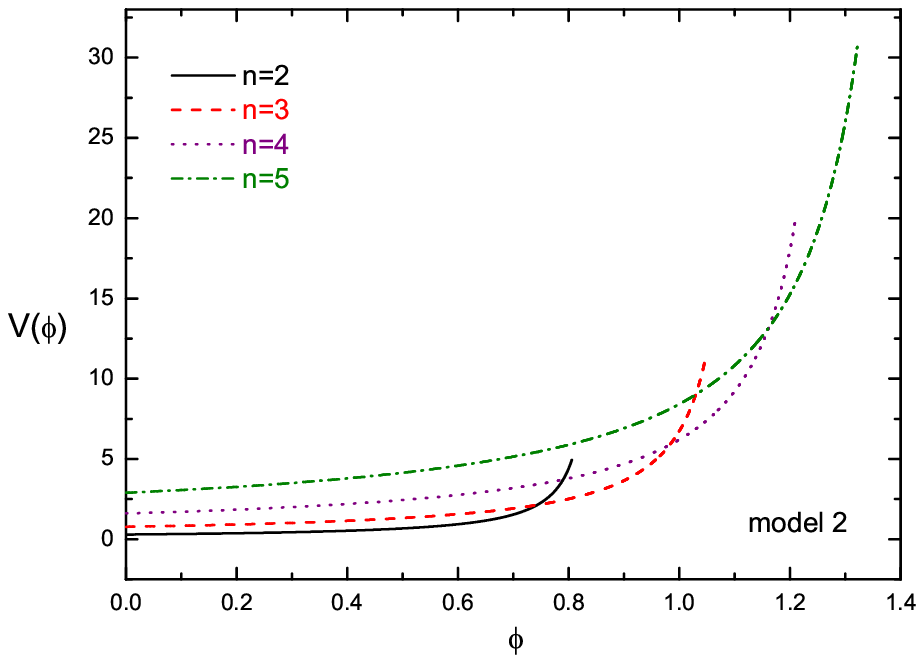} \\
\end{array}$
\caption{The reconstruction of the potential for the agegraphic
quintessence where $\phi$ is in unit of $m_p$ and $V(\phi)$ in
$\rho_{\rm c0}$. We take here $\Omega_{\rm
m0}=0.28$.}\label{fig:vphi}
\end{figure}



\begin{figure}[htbp]
\centering $\begin{array}{c@{\hspace{0.2in}}c}
\multicolumn{1}{l}{\mbox{}} &
\multicolumn{1}{l}{\mbox{}} \\
\includegraphics[scale=0.75]{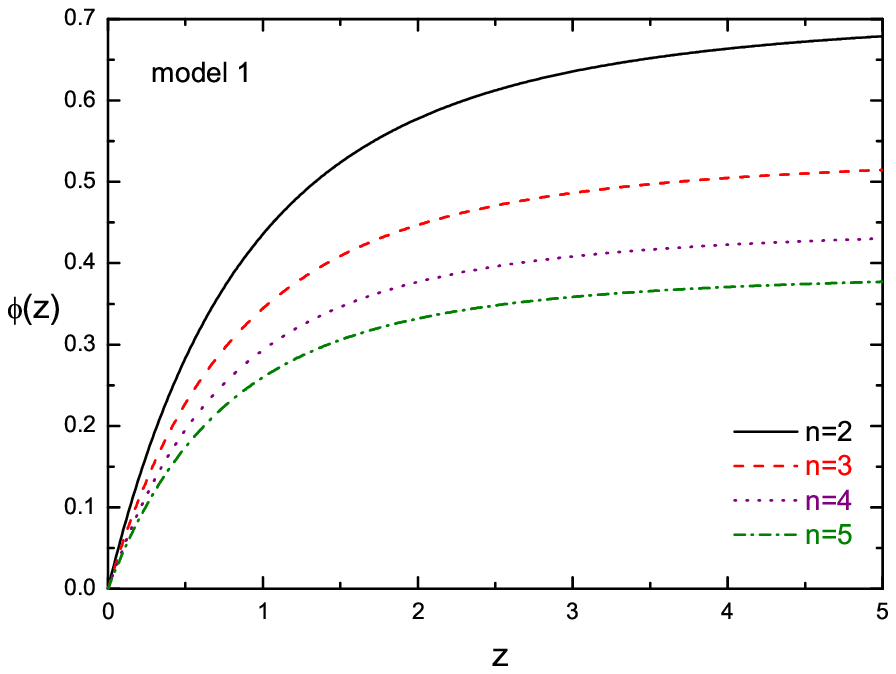} &\includegraphics[scale=0.75]{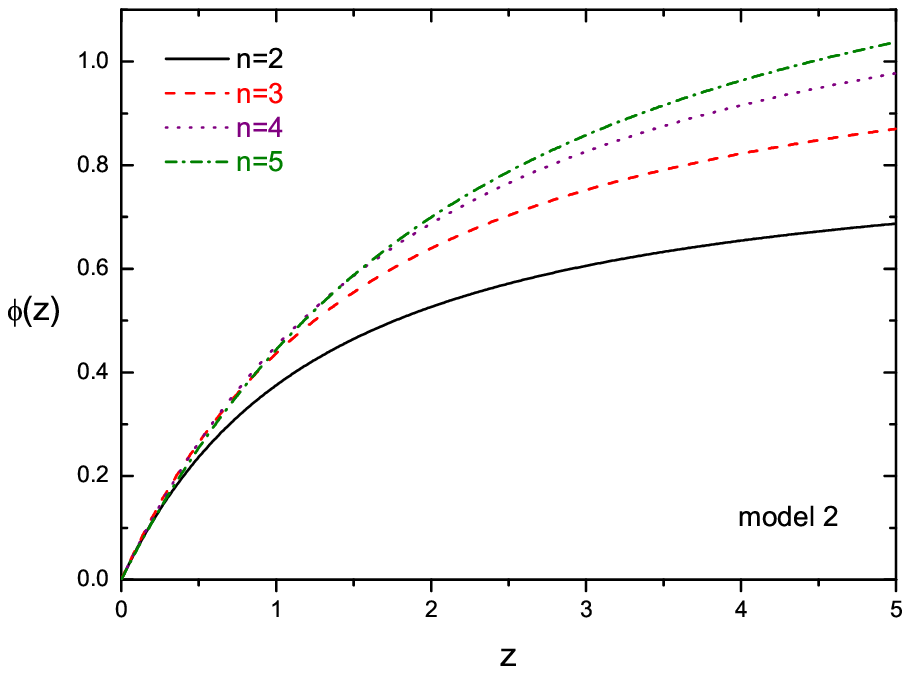} \\
\end{array}$
\caption{The revolution of the scalar-field $\phi(z)$ for the
agegraphic quintessence, where $\phi$ is in unit of $m_p$. We take
here $\Omega_{\rm m0}=0.28$.}\label{fig:phiz}
\end{figure}



 Now we are
focussing on the reconstruction of the agegraphic quintessence. We
shall reconstruct the quintessence potential and the dynamics of the
scalar field in the light of the agegraphic dark energy with $n\geq
1$. According to the forms of quintessence energy density and
pressure, one can easily derive the scalar potential and kinetic
energy term as
\begin{equation}
{V(\phi)\over\rho_{\rm c0}}={1\over 2}(1-w_\phi)\Omega_\phi
E^2,\label{eq18}
\end{equation}
\begin{equation}
{\dot{\phi}^2\over \rho_{\rm c0}}=(1+w_\phi)\Omega_\phi
E^2,\label{eq19}
\end{equation}
where $\rho_{\rm c0}=3m_{\rm p}^2H_0^2$ is today's critical density
of the universe. If we establish the correspondence between the
agegraphic dark energy with $n\geq 1$ and quintessence scalar field,
then $E$, $\Omega_\phi$ and $w_\phi$ are given by Eqs. (\ref{Ez}),
(\ref{eq10}) and (\ref{eq11}) for model 1, while by Eqs. (\ref{Ez}),
(\ref{eq16}) and (\ref{eq17}) for model 2. Furthermore, the
derivative of the scalar field $\phi$ with respect to the redshift
$z$ can be given by
\begin{equation}
{d\phi\over dz}=\pm{m_{\rm p}\sqrt{3(1+w_\phi)\Omega_\phi}\over
1+z},\label{eq20}
\end{equation}
where the sign is actually arbitrary since it can be changed by a
redefinition of the field, $\phi\rightarrow -\phi$. Consequently,
we can easily obtain the evolutionary form of the field
\begin{equation}
\phi(z)=\int\limits_0^z{d\phi\over dz}dz,\label{eq21}
\end{equation}
by fixing the field amplitude at the present epoch ($z=0$) to be
zero, $\phi(0)=0$.


The reconstructed quintessence potential $V(\phi)$ is plotted in
figure \ref{fig:vphi}, where $\phi(z)$ is also reconstructed
according to Eqs. (\ref{eq20}) and (\ref{eq21}), also displayed in
figure \ref{fig:phiz}. Selected curves are plotted for the cases of
$n=2, 3, 4 $ and $5$, and the present fractional matter density is
chosen to be $\Omega_{\rm m0}=0.28$. From figures \ref{fig:vphi} and
\ref{fig:phiz}, we can see the dynamics of the scalar field
explicitly.
As suggested in \cite{Caldwell:2005tm}, quintessence models can be
divided into two classes, ``thawing'' models and ``freezing''
models. Thawing models depict those scalar fields that evolve from
$w= -1$ but grow less negative with time as $dw/d\ln a>0$; freezing
models, whereas, describe those fields evolve from $w>-1$, $dw/d\ln
a<0$ to $w\rightarrow -1$, $dw/d\ln a\rightarrow 0$. Roughly, the
agegraphic quintessence corresponding to model 1 should be ascribed
to the ``thawing'' model, while the agegraphic quintessence
corresponding to model 2 should be ascribed to ``freezing'' model.
As we have seen, the dynamics of the agegraphic quintessence can be
explored explicitly by the reconstruction.


\section{Conclusion}\label{sec5}
In conclusion, we suggest in this paper a correspondence between the
agegraphic dark energy scenario and the quintessence scalar-field
model. We adopt the viewpoint that the scalar field models of dark
energy are effective theories of an underlying theory of dark
energy. A new dark energy model, named as ``agegraphic dark
energy'', has been proposed by Cai \cite{rgcai}, based on the
K\'{a}rolyh\'{a}zy uncertainty relation, which arises from the
quantum mechanics together with general relativity. Perhaps, one may
argue that the agegraphic dark energy lacks convincing physical
foundation and the two models with energy densities in Eqs. (5) and
(12) considered in this paper are just two types of the
phenomenologically parametrized equation of state of dark energy.
However, anyway, one should admit that it is an interesting attempt
to consider the nature of dark energy based on some combination of
quantum mechanics and general relativity. 

If we regard the scalar-field model (such as quintessence) as an
effective description of such a theory, we should be capable of
using the scalar-field model to mimic the evolving behavior of the
agegraphic dark energy and reconstructing this scalar-field model
according to the evolutionary behavior of agegraphic dark energy. We
show that the agegraphic dark energy with $n\geq 1$ can be described
totally by the quintessence in a certain way. A correspondence
between the agegraphic dark energy and quintessence has been
established, and the potential of the agegraphic quintessence and
the dynamics of the field have been reconstructed.


\section*{ACKNOWLEDGMENTS}

This work was supported partially by grants from the National
Natural Science Foundation of China (No. 10573003 and No. 10705041),
the China Postdoctoral Science Foundation (No. 20060400104), the K.
C. Wong Education Foundation and the National Basic Research Program
of China (No. 2003CB716300).

\renewcommand{\baselinestretch}{1.1}


\end{document}